\documentclass[aps,pre,twocolumn,epsfig,floats,superscriptaddress]{revtex4}
\usepackage{amsmath}
\usepackage{color}
\usepackage{graphicx}
\usepackage{amsfonts}
\usepackage{mathrsfs}
\usepackage{amsmath}

\begin{document}

\title{\textbf{Dynamics of quasiperiodically driven spin systems}}

\author{Sayak Ray}
\affiliation{Department of Chemistry, Ben-Gurion University of the Negev, Beer-Sheva 84105, Israel}

\author{Subhasis Sinha}
\affiliation{Indian Institute of Science Education and Research,
Kolkata, Mohanpur, Nadia 741246, India}

\author{Diptiman Sen}
\affiliation{Centre for High Energy Physics, Indian Institute of Science, 
Bengaluru 560012, India}

\date{\today}
\begin{abstract}
We study the stroboscopic dynamics of a spin-$S$ object subjected to 
$\delta$-function kicking in the transverse magnetic field which is generated following the Fibonacci sequence. The corresponding classical Hamiltonian map is constructed in the large spin limit, $S \rightarrow \infty$. Upon evolving such 
a map for large kicking strength and time period, the phase space appears 
to be chaotic; interestingly, however, the geodesic distance increases 
linearly with the stroboscopic time implying that the Lyapunov exponent is zero. We derive the Sutherland invariant for the underlying $SO(3)$ matrix governing the dynamics of classical spin variables and study the orbits for weak kicking strength. For the quantum dynamics, we 
observe that although the phase coherence of a state is retained throughout the time evolution, the fluctuations in the mean values of the spin operators exhibit fractality which is also present in the Floquet eigenstates. Interestingly, the presence of an interaction with another spin results in an ergodic dynamics leading to infinite temperature thermalization.
\end{abstract}

\maketitle

\section{Introduction}

In recent years quasiperiodic systems have attracted a lot of interest in 
various contexts ranging from quasicrystals~\cite{quasi_lattice,q_crys_book,
q_crys,q_crys1,q_crys2} to localization-delocalization transitions 
\cite{Inguscio,hanggi,Huse,Bloch,Bloch1}, multifractality~\cite{mult_frac,
kohmoto,mult_frac1,frac}, topological phases~\cite{q_topo}, and so on. 
The creation of a quasiperiodic potential in one dimension
using bichromatic optical lattices~\cite{Inguscio} has led to the realization of the well known Aubry-Andr\'e model~\cite{AA} which has been studied extensively both theoretically~\cite{hanggi,Huse} and experimentally~\cite{Bloch,Bloch1,AA_light}, in the context of observing localization phenomena, particularly 
many-body localization in interacting systems~\cite{MBL_rev,abanin_rmp}. In the presence 
of a periodic drive such many-body localized states exhibit drive-induced delocalization and thermalization of isolated interacting quantum 
systems~\cite{Bloch_drive,abanin_drive}, and its connection with the underlying chaotic dynamics and random matrix theory has been explored~\cite{Sray_drive}. As an extension, several interesting questions can be addressed related to the dynamical behavior of quantum systems under a quasiperiodic drive.
One such issue is the emergence of steady states in quasiperiodically 
driven interacting quantum systems~\cite{a_sen,a_dutta} and most 
interestingly, its connection with spectral properties and random matrix 
theory which is related to ergodicity.

One way to generate a quasiperiodic drive is by perturbing the system under 
consideration following the Fibonacci sequence; such a sequence has a rich 
mathematical structure giving rise to an invariant of the corresponding 
dynamical systems~\cite{pandit_inv,kohmoto_inv,Sutherland}.
This way of generating a quasiperiodic drive can provide an alternate way 
to study the quasiperiodic structures observed in Fibonacci 
lattices~\cite{kohmoto_inv,Sutherland,Ksaha}. The Fibonacci drive can also 
be generated for a series of incommensurate frequencies known as metallic 
means, a common example of which is the golden ratio, $\beta=(\sqrt{5}+1)/2$.
These driving protocols can give rise to the realization of a strange non-chaotic attractor~\cite{Ruelle} leading to a fractal-like dynamics which has been theoretically studied for dynamical maps with 
quasiperiodicity~\cite{fractal,pikovsky,lai,ramaswamy} as well observed 
experimentally~\cite{SNA_exp}. The evolution of a spin-$1/2$ system under 
quasiperiodic perturbation reveals various interesting dynamical behaviors and 
temporal correlations~\cite{orland,lebowitz,geisel_quasi}.
A quasiperiodic drive can also lead to slow relaxation to non-equilibrium 
steady states which has been investigated for interacting spin systems in 
the presence of a disordered magnetic field~\cite{potter_mag}. 
It is interesting to investigate if there is any underlying fractality in 
the eigenmodes of the time evolution operator even for non-interacting 
systems under a quasiperiodic drive and the fate of such 
critical states in the presence of interactions.

\begin{figure}[ht]
\centering
\includegraphics[scale=0.43]{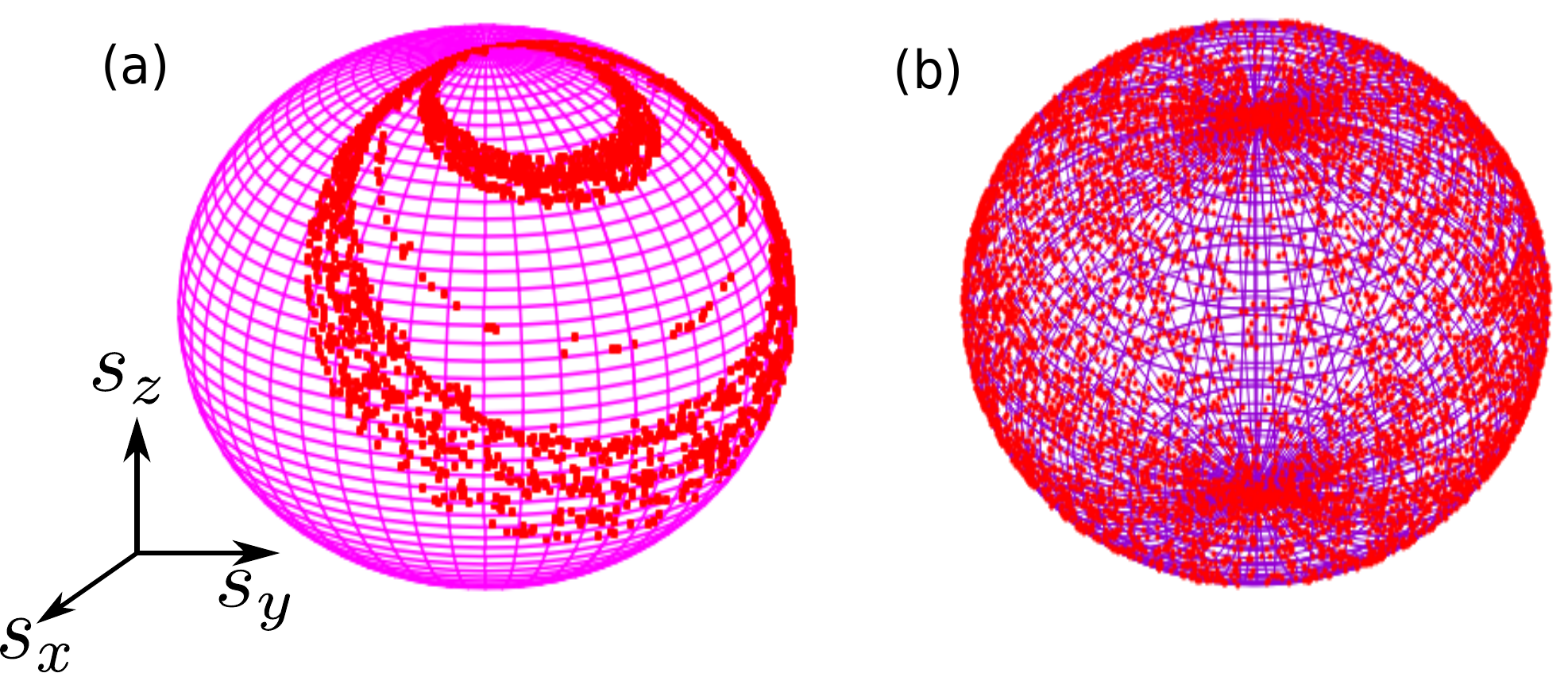}
\caption{Stroboscopic dynamics of the spin variables $s_x$, $s_y$ and $s_z$ on 
a unit sphere for (a) $\lambda=\pi/100$, $T_0=\pi/50$ and (b) $\lambda=\pi/10$,
$T_0=\pi/10$.} \label{pp_spin} 
\end{figure}

In this work we first consider a simple model of a non-interacting spin of magnitude $S$ (which has a well-defined classical limit) subjected to a quasiperiodic drive. Our objective is two-fold. First, we want to study the dynamics of the corresponding classical system and analyze the fluctuations around the mean values of the spin components.
Second, we want to study the quantum dynamics and spectral properties of the time
evolution operator, particularly its change in the presence of interaction.
The paper is organized as follows. In Sec.\ \ref{Sec1} we describe the 
time-dependent Hamiltonian and the details of the quasiperiodic driving 
protocol. We then study the dynamics of the corresponding classical system by 
suitably taking the limit $S \rightarrow \infty$ in Sec.\ \ref{class_dyn}. 
This is followed by the derivation of an invariant for this dynamical system 
in Sec.\ \ref{Sutherland}. The quantum dynamics of a finite spin $S$ and the 
spectral properties of the time evolution operator are discussed in 
Sec.\ \ref{qdyn}. The effects of interaction are studied in 
Sec.\ \ref{interaction}. Finally, we summarize our results, discuss possible 
experiments which can test our results, and conclude in Sec.\ \ref{conclu}.

\section{Driven spin model in transverse magnetic field}
\label{Sec1}

The dynamics of a spin-$S$ particle in the presence of a time-dependent magnetic field can be described by the Hamiltonian
\begin{equation} 
\hat{H}(t) ~=~ \omega_0 \hat{S}_z ~+~ \lambda \hat{S}_x 
\sum_{n=-\infty}^{\infty} \delta \left(t-\sum_n T_n \right), \label{Eq1} 
\end{equation}
where the first term in the Hamiltonian represents the effect of a
magnetic field with strength $\omega_0$ applied along the $\hat z$ direction
(we have absorbed the gyromagnetic ratio in the definition of $\omega_0$), and
the second term represents a $\delta$-function 
kicking due to a transverse magnetic field in the 
$\hat x$ direction with strength $\lambda$. Here $\hat{S}_{x,y,z}$'s denote the 
spin angular momentum operators and $T_n$ is the time lapse between the 
$(n-1)$-th and $n$-th kicks. Here we consider the case $T_n = T_0 (1 \pm \epsilon) \equiv T_{\pm}$, i.e., the time lapse can take two values $T_{+}$ or $T_{-}$ 
which follows the Fibonacci sequence:
\begin{equation}
\text{$T_{+}$, $~T_{-}$, $~T_{+}$, $~T_{+}$, $~T_{-}$, $~T_{+}$, $~T_{-}$, 
$~T_{+}$, $~\cdots$}. \label{Fibo_seq} \end{equation}
We will scale the energy (time) by $\omega_0$ $(1/\omega_0)$ and set 
$\hbar = 1$ throughout this paper.

The time evolution operator describing the dynamics of such system between 
the $(n-1)$-th and $n$-th kicks is given by
\begin{equation}
\hat{\mathcal{F}}_n ~=~ e^{-iT_n\hat{S}_z} ~e^{-i\lambda \hat{S}_x}.
\label{Floquet} 
\end{equation}
Note that for a periodic drive in which all the $T_n$'s are equal, 
Eq.\ \eqref{Floquet} reduces to the usual Floquet operator. 
Following the time evolution of any operator between the $(n-1)$-th and $n$-th kicks 
under the Floquet matrix $\hat{\mathcal{F}}_n$ (namely, $\hat{A}_{n+1} = 
\hat{\mathcal{F}}_n^{\dagger} \hat{A}_{n} \hat{\mathcal{F}}_n$) 
\cite{Floquet}, we obtain the Heisenberg equations of motion for the spin operators as follows,
\begin{eqnarray} \hat{S}_x^{n+1} &=& \hat{S}_x^{n} \cos T_n - \sin T_n 
(\hat{S}_y^{n} \cos \lambda - \hat{S}_z^{n} \sin \lambda), \nonumber \\
\hat{S}_y^{n+1} &=& \hat{S}_x^{n} \sin T_n + \cos T_n (\hat{S}_y^{n} \cos 
\lambda - \hat{S}_z^{n} \sin \lambda), \nonumber \\
\hat{S}_z^{n+1} &=& \hat{S}_y^{n} \sin \lambda + \hat{S}_z^{n} \cos \lambda. \label{heisen_eq}
\end{eqnarray} 
This is a linear map for the spin operators of the form: $(\hat{S}_x^{n+1}, \hat{S}_y^{n+1}, \hat{S}_x^{n+1}) = J_n (\hat{S}_x^{n}, \hat{S}_y^{n}, \hat{S}_x^{n})$, where the transfer matrix $J_n$ can be written as
\begin{equation} J_{n} ~=~ \left( \begin{array}{ccc}
\cos T_{n} & -\sin T_{n} ~\cos \lambda & \sin T_{n} ~\sin \lambda \\
\sin T_{n} & \cos T_{n} ~\cos \lambda & -\cos T_{n} ~\sin \lambda \\
0 & \sin \lambda & \cos \lambda, \end{array}\right), \label{tran_mat} 
\end{equation}
where the $T_n$'s are given in Eq.\ \eqref{Fibo_seq}. 
We will consider the case $\epsilon = 1$, so that Eq.\ \eqref{Floquet} becomes
\begin{equation}
\hat{\mathcal{F}}_1 ~=~ e^{-i\lambda \hat{S}_x}, \quad \quad
\hat{\mathcal{F}}_2 ~=~ e^{-i2T_0\hat{S}_z}e^{-i\lambda \hat{S}_x},
\label{fibo1} \end{equation}
where $\hat{\mathcal{F}}_1$ represents kicking the spin-S object by a 
magnetic field in the $\hat x$ direction, and $\hat{\mathcal{F}}_2$ corresponds 
to time evolution of the system under $\hat{S}_z$ for time interval
$2T_0$ followed by another kick in $\hat{S}_x$. 

Starting from two such $SU(N)$ matrices, $\hat{\mathcal{F}}_1$ and 
$\hat{\mathcal{F}}_2$, the successive Floquet operators in a Fibonacci 
sequence can be generated using the recursion relation, 
\begin{equation}
\hat{\mathcal{F}}_{m+2} ~=~ \hat{\mathcal{F}}_{m+1} \hat{\mathcal{F}}_{m},
\label{fibo_rec} 
\end{equation}
where the initial matrices $\hat{\mathcal{F}}_1$ and $\hat{\mathcal{F}}_2$ 
are given in Eq.\ \eqref{fibo1}. We would like to point out that at a 
Fibonacci time step $m$, the stroboscopic time is given by $n=F_{m}$, where 
$F_{m}$ is the $m$-th Fibonacci number. For large $m$, the stroboscopic time 
increases exponentially as $n \sim e^{\beta_G m}$ where, 
$\beta_G = (\sqrt{5}+1)/2$ is the golden ratio. The advantage of using such 
a recursion relation is that one can numerically obtain the steady state of 
the system after a very long time scale. Henceforth, we will adopt 
Eq.\ \eqref{fibo_rec} to study the dynamics for a very large duration in 
stroboscopic time. 

\subsection{Classical Dynamics}
\label{class_dyn}

\begin{figure}[ht]
\centering
\includegraphics[scale=0.17]{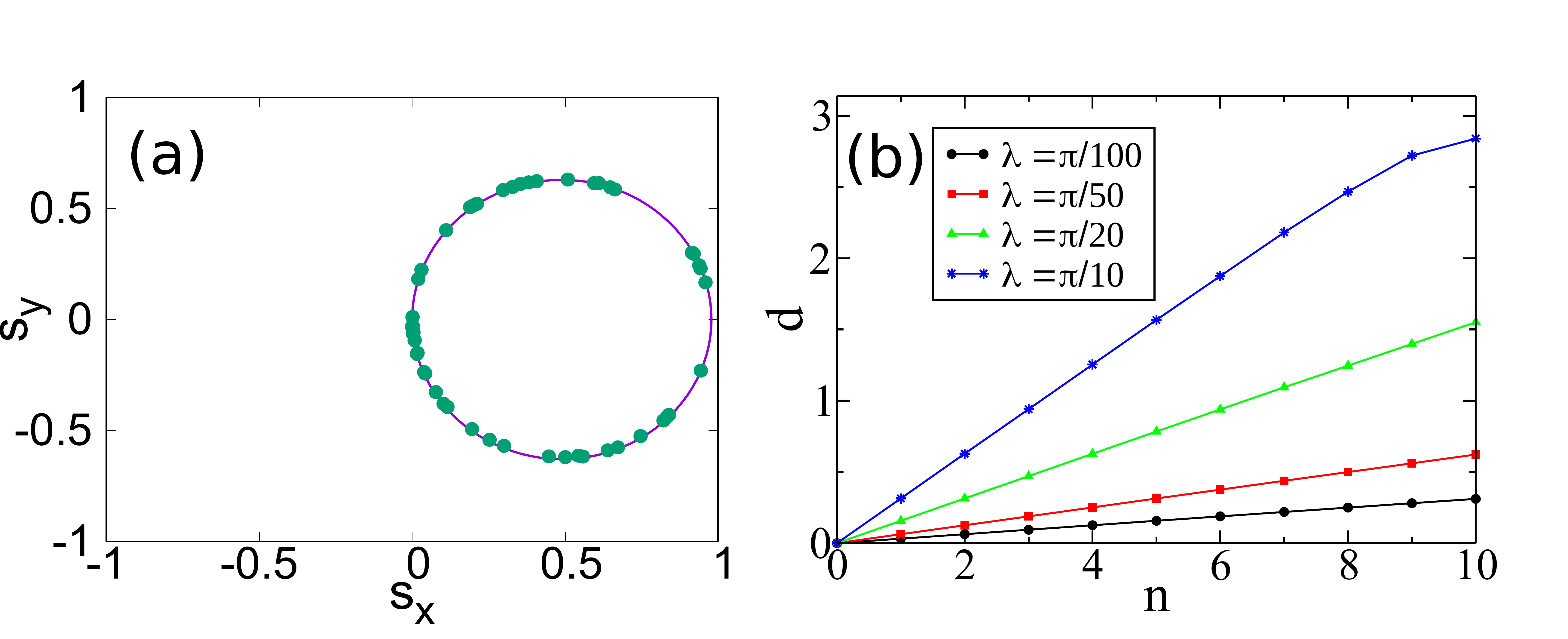}
\caption{(a) Dynamics of the spin variables projected onto the $s_x-s_y$ plane 
for $T_0 = \pi/100$ and $\lambda = \pi/100$ starting from initial point $(0,0,1)$. 
The dotted points are obtained 
numerically and the solid line is drawn using an analytical expression. (b) 
Geodesic distance $d$ as a function of stroboscopic time $n$.} 
\label{lyap} 
\end{figure}

We will first discuss the dynamics of the corresponding classical system. 
The classical limit of such a spin system can be obtained by considering the 
large spin limit, $S \rightarrow \infty$. Then the spin variables $\hat{S}_{x,y,z}$ can be classically described by the components of a spin vector $\vec{S} \equiv (S_x,S_y,S_z)$. We scale the spin operators by the magnitude S to obtain the classical spin variables,
$s_i = S_i/S$ which follow the commutation relations $[s_i,s_j]=i\epsilon_{ijk}s_k/S$. In the limit $S \rightarrow \infty$, the commutators vanish and the 
variables become classical. Thus using Eqs.\ \eqref{heisen_eq}, the 
stroboscopic time evolution of the corresponding classical spin variables 
in between consecutive kicks can be described by the following linear 
Hamiltonian map,
\begin{equation} \left( \begin{array}{c}
s_x^{n+1} \\
s_y^{n+1} \\
s_z^{n+1} \end{array}\right)
~=~ J_n ~\left( \begin{array}{c}
s_x^{n} \\
s_y^{n} \\
s_z^{n} \end{array}\right), \label{ham_map} 
\end{equation}
where the transfer matrix $J_n$ is given in Eq.\ \eqref{tran_mat}. By evolving Eq.\ \eqref{ham_map} stroboscopically in time we obtain the trajectories on a unit sphere as shown in Fig.\ \ref{pp_spin} for different driving parameters. We observe that for a small driving strength $\lambda$ and time period $T_0$ the trajectories are regular and precess over time as depicted in Fig.\ \ref{pp_spin} (a). Over a small time scale such a regular trajectory is plotted in the 
projected plane of $s_x-s_y$ in Fig.\ \ref{lyap} (a). However, for large 
values of $\lambda$ and $T_0$, the dynamics is no longer regular and eventually
covers the whole surface of the sphere shown in Fig.\ \ref{pp_spin} (b).
It is interesting to note that the transfer matrix in Eq.\ \eqref{tran_mat} is unimodular and therefore its eigenvalues have the form: $1,e^{\pm i\varepsilon}$, $\varepsilon$ being the eigenphase. As a result, the
Lyapunov exponent always turns out to be zero~\cite{blumel,schack,kurths}. To further illustrate this, we compute the growth of the
geodesic distance on the Bloch sphere. We start from the initial point $s_i = (s_x,s_y,s_z) = (0,0,1)$ and evolve it under successive kicking following Fibonacci sequence. The resulting trajectory
in $s_x-s_y$ plane is depicted in Fig.\ \ref{lyap}a. The geodesic distance 
between the initial point $s_i$ and the time-evolved point $s_f$ is given
by $d = \cos^{-1}({\vec s}_i \cdot {\vec s}_f)$. In Fig.\ \ref{lyap}b we have 
plotted $d$ as a function of the stroboscopic time $n$; we see a linear growth which 
indicates that the Lyapunov exponent is zero in this case.

\subsection{Fibonacci sequence of $SO(3)$ matrices and Sutherland invariant}
\label{Sutherland}

We first discuss the case of $SU(2)$ matrices multiplied according 
to a Fibonacci sequence for which Sutherland found an invariant; then we will 
discuss how this invariant generalizes to the case of $SO(3)$ matrices. 
Starting with two $SU(2)$ matrices $U_1$ and $U_2$, we generate a Fibonacci sequence of matrices defined by the recursion relation $U_{m+2}=U_{m+1}U_m$. Let us parametrize $U_m$ as,
\begin{equation} U_m ~=~ e^{i\alpha_m\hat{n}_m \cdot \vec{\sigma}} ~=~ \cos 
\alpha_m \,\, \mathbb{I}_2 ~+~ i ~\sin \alpha_m \,\, \hat{n}_m \cdot 
\vec{\sigma}, 
\end{equation}
where $0 \leq \alpha_m \leq \pi$, $\hat{n}_m$ is a unit vector, $\mathbb{I}_2$ 
denotes the $2\times 2$ identity matrix, and $\vec{\sigma}=(\sigma_x,\sigma_y,\sigma_z)$ denotes the Pauli matrices. Defining $x_m=\frac{1}{2}\text{tr}(U_m)
= \cos \alpha_m$, Sutherland showed that the quantity
\begin{equation} I_s ~=~ x_m^2 ~+~ x_{m+1}^2 ~+~ x_{m+2}^2 ~-~ 
2x_m x_{m+1} x_{m+2} ~-~ 1 \label{is} 
\end{equation} 
is independent of $m$~\cite{Sutherland}. 

We now consider $SO(3)$ matrices denoted as $R_m$. Starting with two such 
matrices $R_1$ and $R_2$, we generate a Fibonacci sequence using the 
recursion relation
\begin{equation} R_{m+2} = R_{m+1}R_m. \end{equation}
Let us parametrize $R_m$ as follows,
\begin{equation} R_m ~=~ e^{i\phi_m\hat{e}_m \cdot \vec{T}}, \end{equation}
where $0 \leq \phi_m \leq 2\pi$, $\hat{e}_m$ is a unit vector 
and $\vec{T}=(T_x,T_y,T_z)$ are the generators of $SO(3)$ matrices. 
One can show that the matrix elements of $R_m$ given by $R^m_{ij}$ 
and the components of $\hat{e}_m$ given by $e_i^m$ are related as
\begin{eqnarray} R^m_{ij} &=& \delta_{ij} \cos \phi_m ~+~ e^m_i e^m_j 
(1-\cos \phi_m) \nonumber \\
&& + ~\sum_{k=1}^3 ~\epsilon_{ijk} ~e^m_k \sin \phi_m, \label{Rm} \end{eqnarray}
where $\epsilon_{ijk}$ is the totally antisymmetric matrix with 
$\epsilon_{123} = 1$. Now using the standard mapping between the spin-1/2 and spin-1 representations of the angular momentum group, $\alpha_m=\phi_m/2$ and 
$\hat{n}_m=\hat{e}_m$ one can obtain the Sutherland invariant $I_s$ for a given
$SO(3)$ matrix $R_m$. [There is a subtlety here: since the same $SO(3)$
matrix $R_m$ corresponds to two different $SU(2)$ matrices, $U_m$ and $-U_m$,
whose traces (divided by 2) are given by $x_m$ and $-x_m$, one has to check 
at each step of the Fibonacci sequence which of the two possible values of the 
trace gives the correct value of the Sutherland invariant in Eq.~\eqref{is}.]

As an example we consider the transfer matrix $J_n$, given in 
Eq.\ \eqref{tran_mat} which is a $SO(3)$ matrix. Starting with two such 
matrices $J_1$ and $J_2$ given by
\begin{equation} J_1 ~=~ e^{-i\lambda T_x} \quad \text{and} \quad J_2 ~=~
e^{-i2T_0T_z}e^{-i\lambda T_x}, \end{equation}
we find that the corresponding Sutherland invariant is
\begin{equation} I_s ~=~ - ~[\sin T_0 \, \sin (\lambda/2)]^2. \label{isuth} 
\end{equation}
We have checked numerically up to a large Fibonacci step $m \sim 1000$ that 
$I_s$ remains constant during the time evolution. 

It is known that the $m$-th Fibonacci number, given by $F_m=\left(\beta_G^m-
\frac{(-1)^m}{\beta_G^m}\right)/\sqrt{5}$, quickly approaches the value 
$\beta_G^m/\sqrt{5}$ as $m$ increases. We then find that
\begin{equation} J_m ~=~ e^{-i\beta_G^m[\lambda T_x+(2T_0/\beta_G)T_z]/
\sqrt{5}}, \end{equation}
which has been derived under the approximation that $\lambda$, $T_0 \ll 1$ 
so that the commutators arising from $[T_i,T_j]$ do not grow much within 
a small time scale. This leads us to define
\begin{equation} \phi_m = -\frac{\beta_G^m}{\sqrt{5}}\bar{\lambda} \quad
\text{and} \quad \hat{e}_m = \frac{1}{\bar{\lambda}}(\lambda,0,2T_0/\beta_G), 
\end{equation}
where $\bar{\lambda}=\sqrt{\lambda^2+4T_0^2/\beta_G^2}$. We will now study what happens when $R_m$ acts on the column $(s_x,s_y,s_z)=(0,0,1)$ as numerically 
shown in Fig.\ \ref{lyap} (a). Using Eq.\ \eqref{Rm}, we see that
\begin{equation} \left( \begin{array}{c}
s_x^{m} \\
s_y^{m} \\
s_z^{m} \end{array}\right)
=\left( \begin{array}{c}
e_1e_3(1-\cos \phi_m)-e_2\sin \phi_m \\
e_2e_3(1-\cos \phi_m)+e_1\sin \phi_m \\
\cos \phi_m+e_3^2(1-\cos \phi_m) \end{array}\right). \end{equation}
We then see that
\begin{equation} (s_x^m,s_y^m) ~=~ (e_1e_3(1-\cos \phi_m),e_1\sin \phi_m),
\end{equation}
where $e_1=\lambda/\bar{\lambda}$ and $e_3=2T_0/\beta_G \bar{\lambda}$. We see 
that the point $(s_x^m,s_y^m)$ describes an ellipse as shown by the solid line 
in Fig.\ \ref{lyap} (a) whose center lies at $(e_1e_3,0)$ and the lengths of 
the axes are $2e_1e_3$ and $2e_1$ in the $\hat x$ and $\hat y$ directions respectively.

We would like to mention here that quasiperiodic driving of $SU(2)$ matrices 
has been studied in detail in Refs.~\cite{a_sen} and~\cite{a_dutta}.
It has been found that the long-time behavior of the system depends to some
extent on the value of the Sutherland invariant $I_s$~\cite{a_sen}. The 
behavior is particularly 
simple near $I_s = 0$ and $-1$ (which are respectively the maximum and minimum 
possible values of $I_s$). Near $I_s = 0$, the trajectory
is given by a circle on the Bloch sphere~\cite{a_dutta}. This is similar to 
the behavior of $SO(3)$ matrices discussed above, namely, the point moves on 
an ellipse when $T_0$ and $\lambda$ are small which corresponds to a very 
small value of $I_s$ according to Eq.\ \eqref{isuth}.

\subsection{Quantum Dynamics}
\label{qdyn}

We now analyze the quantum dynamics of a spin-$S$ object governed by the 
time-dependent Hamiltonian in Eq.\ \eqref{Eq1}. To this end we construct the 
initial wave function $|\psi(0)\rangle$ from a spin coherent state given 
by~\cite{radcliff}
\begin{equation} |\Theta, \Phi\rangle ~=~ (1+|z|^2)^{-S}e^{z\hat{S}_{+}} ~ 
|S,-S\rangle, 
\end{equation}
where $\Theta$ and $\Phi$ are the polar and azimuthal angles, respectively, 
representing the orientation of the classical spin vector of magnitude $S$, and 
$z = e^{-i\Phi}\tan (\Theta/2)$. The time-evolved state after the $m$-th
Fibonacci kick is given by
\begin{equation} 
|\psi(m)\rangle ~=~ \hat{\mathcal{F}}_m ~ |\psi(0)\rangle.
\end{equation}

\begin{figure}[ht]
\centering
\includegraphics[scale=0.16]{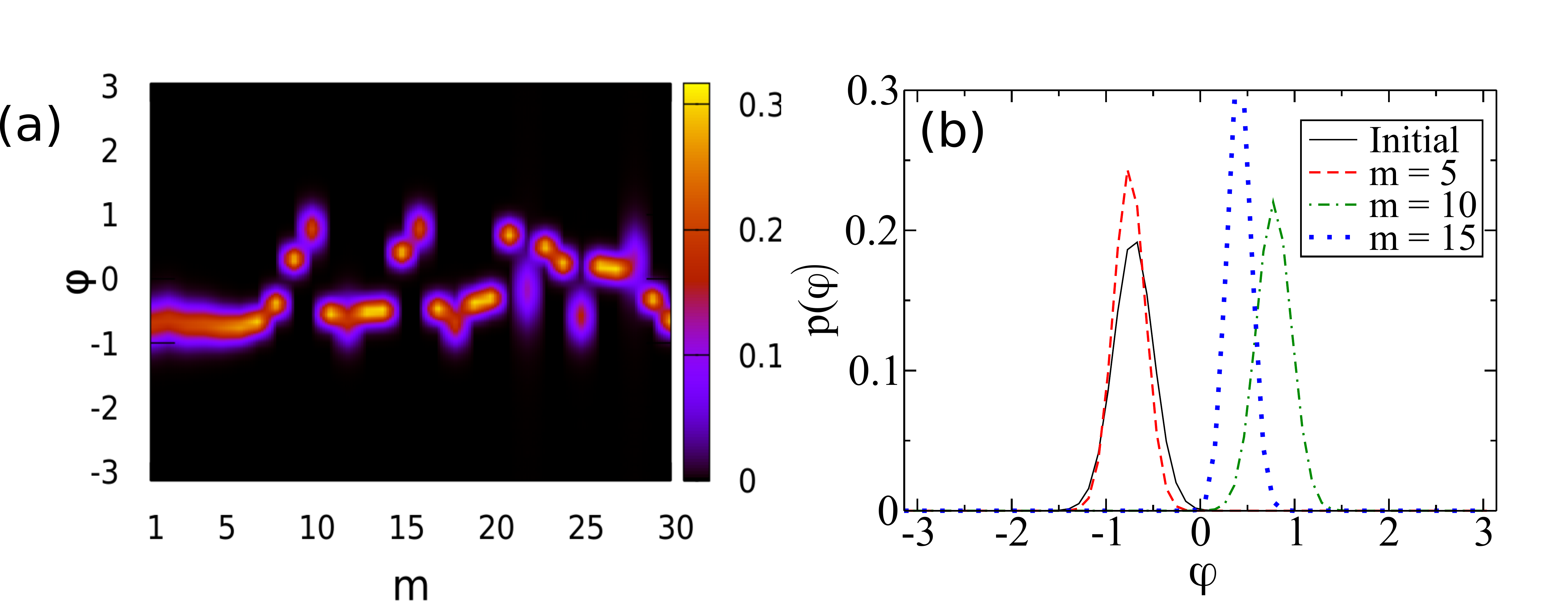}
\caption{(a) Time evolution of $p(\varphi)$. (b) Snapshots of $p(\varphi)$ at different Fibonacci time step for $T_0 = \pi/100$ and $\lambda = \pi/40$.}
\label{p_diff}
\end{figure}

To compute the distribution of the relative phases of the time-evolved state 
we first construct the phase state given by~\cite{oberthaler},
\begin{equation} |\varphi\rangle ~=~ \frac{1}{\sqrt{2S+1}}\sum_{l=1}^{2S+1}
e^{il\varphi} |l\rangle, \end{equation}
where $\varphi = \varphi_0 + 2\pi l^{\prime}/(2S+1)$ and 
$l^{\prime} \in [1,2S+1]$. We choose $\varphi_0 = -\pi$ so that the relative 
phase lies in the range $-\pi$ to $\pi$. The phase distribution can be obtained by projecting $|\psi(m) \rangle$ onto the phase state as given by
\begin{equation} 
p(\varphi) = |\langle \varphi| \psi(m)\rangle|^2. 
\end{equation}
In Fig.\ \ref{p_diff} (a) we have shown the time evolution of $p(\varphi)$, 
and in Fig.\ \ref{p_diff} (b) the snapshots of $p(\varphi)$ at different times 
are plotted. We observe the phase distribution $p(\varphi)$ remains a highly peaked function, 
however, its peak position changes under stroboscopic evolution. 
It indicates that the phase diffusion does not take place even for 
large kicking strengths, and the phase coherence is not lost during the time evolution.
We also compute the expectation values of the spin operators, i.e., 
$\langle \hat{S}_z\rangle$ where the average $\langle . \rangle$ is taken with 
respect to the time-evolved state $|\psi(m)\rangle$. We have verified that under quantum dynamics $\langle \hat{S}_i \rangle$'s are in agreement with the classical variables $s_i$ obtained from the dynamical map.
The time evolution of the mean values of these operators exhibit 
fluctuations similar to the peak of $p(\varphi)$ in Fig.\ \ref{p_diff}. 

To understand the nature of the fluctuations in the dynamics of spin variables,
we compute the cumulative sum of the Fourier transform of $\langle \hat{S}_z
\rangle$ given by~\cite{pikovsky,lai,ramaswamy,orland,lebowitz},
\begin{equation} 
X_{\Omega} ~=~ \sum_{m=1}^{N} x_m e^{i2\pi \Omega m},
\end{equation}
where $x_m = \langle \hat{S}_z\rangle$ computed at the $m$-th Fibonacci step, 
and $\Omega$ is the frequency. From the power spectrum, $X_{\Omega}$ vs 
$\Omega$, we observe that several frequency modes are present in the fluctuations which confirms that the dynamics is not at all regular even for small values 
of $\lambda$ and $T_0$ for which the spin variables are seen to
undergo a precessional motion in Fig.\ \ref{pp_spin} (a).
\begin{figure}[ht]
\centering
\includegraphics[scale=0.2]{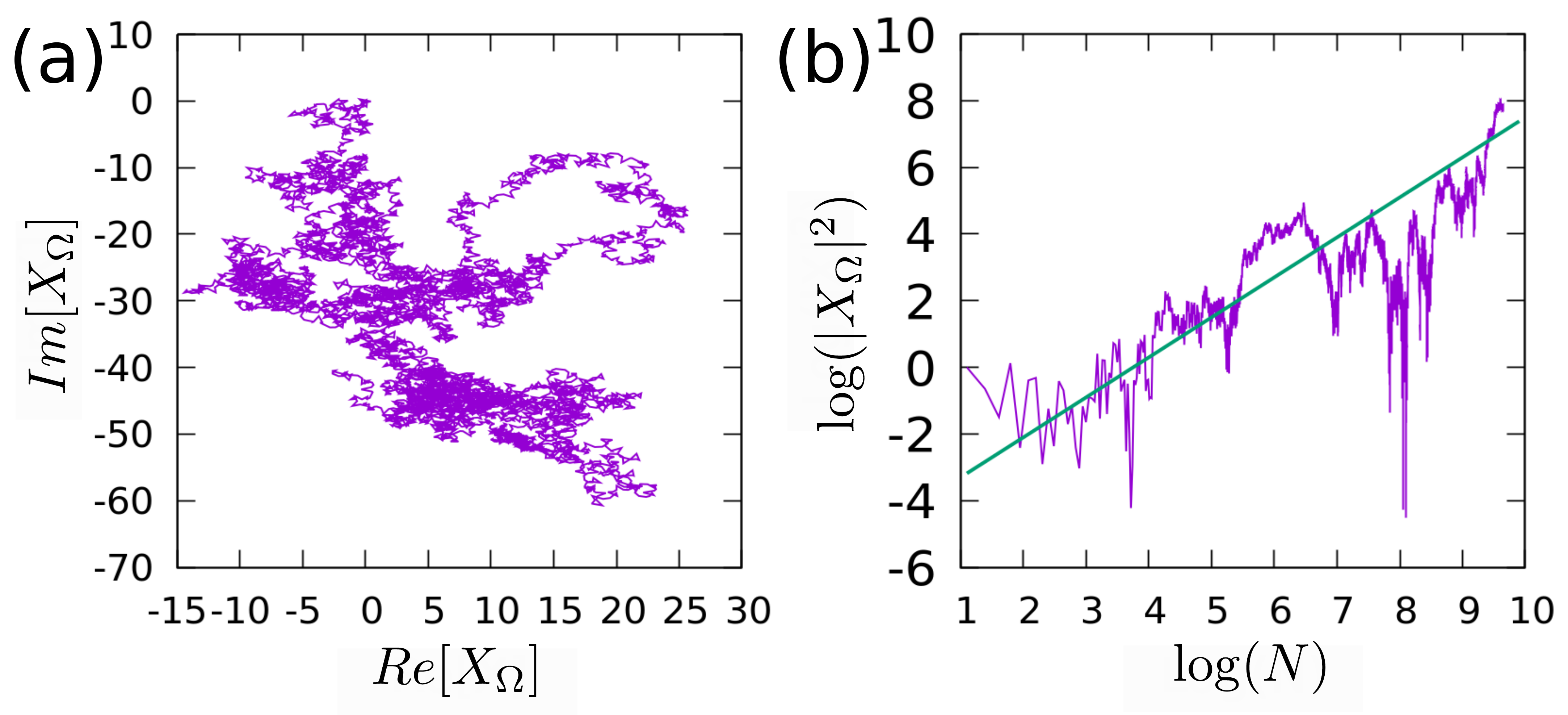}
\caption{(a) Stroboscopic dynamics in the $\text{Re}[X_{\Omega}]-\text{Im}
[X_{\Omega}]$ plane for driving parameters $T_0 = \pi/100$, $\lambda = \pi/50$
and frequency $\Omega = 1/\beta_G$. (b) $|X_{\Omega}|^2$ as a function of the
Fibonacci step $N$. We find that $|X_{\Omega}|^2 \sim N^{1.2}$.}
\label{fractal} 
\end{figure}
The corresponding dynamics in the
$\text{Re}[X_{\Omega}]-\text{Im}[X_{\Omega}]$ plane exhibits a fractal-like structure as depicted in Fig.\ \ref{fractal} (a). The degree of such fractal motion can be quantified using the relation $|X_{\Omega}|^2 \sim N^{\beta}$, where 
the exponent $\beta = 1\, (2)$ signifies random (regular) paths respectively, whereas $\beta \neq 1, 2$ corresponds to a fractal path in the $\text{Re}[X_{\Omega}]-\text{Im}[X_{\Omega}]$ plane which can be observed from the logarithmic plot in Fig.\ \ref{fractal}b. We have computed $\beta$ for different values of $\Omega$, i.e., $\beta(\Omega=0.2)=1.436$ and $\beta(\Omega=0.4)=1.1$, for 
$\lambda = \pi/50$ and $T_0=\pi/100$. We have checked that for different 
choices of $\lambda$ and $T_0$ the behavior remains qualitatively similar. 

\begin{figure}[ht]
\centering
\includegraphics[scale=0.17]{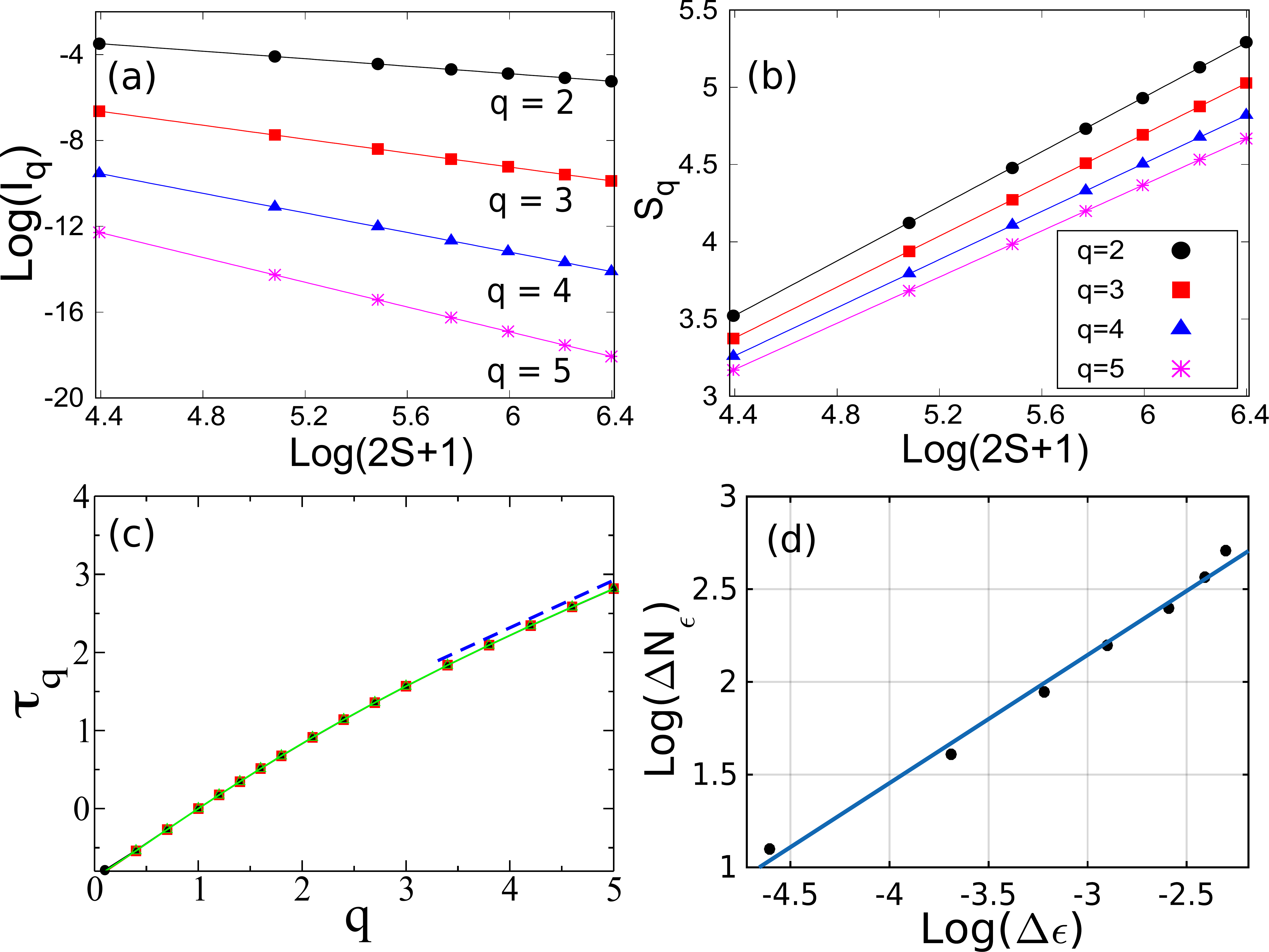}
\caption{(a) $\log I_q$ and (b) $S_q$ are plotted as a function of 
$\log(2S+1)$ for a typical choice of parameters $\lambda = \pi/10$ and 
$T_0 = \pi/10$. (c) $\tau_q$ vs $q$ calculated after a sufficient number of 
Fibonacci steps $m$. Note that the graphs for different $m$ overlap with each 
other. The asymptotic slope (see the dashed line) is $D_q \sim 0.6$. 
(b) $\log \Delta N_{\varepsilon}$ vs $\log \Delta \varepsilon$ with 
$T_0=\pi/100$ and $\lambda=\pi/50$ for $S=400$ computed in the middle of the 
band ($\varepsilon \sim 0$). The dots are numerical data and the solid line 
is fitted with a slope $\alpha \sim 0.69$.} \label{tq_vs_q} \end{figure}

Next, we investigate the spectral properties of the Floquet operator, $\hat{\mathcal{F}}_m$, after a sufficiently large Fibonacci step $m$. Due to unitarity, we have
$\hat{\mathcal{F}}_m |\chi_{\nu}\rangle = e^{i\varepsilon_{\nu}} 
|\chi_{\nu}\rangle$, where $\varepsilon_{\nu} \in [-\pi,\pi]$ and $|\chi_{\nu}\rangle$ are the eigenphase and eigenvector corresponding to the 
$\nu$-th Floquet eigenmode. We compute the moments of the eigenstates 
$|\chi_{\nu}\rangle$ of the Floquet operator $\hat{\mathcal{F}}_m$ at the 
$m$-th Fibonacci step using the relation given by
\begin{equation} 
I_q^{\nu} = \sum_{m=1}^{2S+1} |\chi_{\nu}(m)|^{2q}, 
\quad I_q = \frac{1}{2S+1} \sum_{\nu} I_q^{\nu} \sim (2S+1)^{-\tau_q},
\label{Iq_moment} 
\end{equation}
where $\chi_{\nu}(m) = \langle \chi_{\nu}|\alpha_{m}\rangle$, $|\alpha_m
\rangle$ being the computational basis. The exponent $\tau_q$ is related to 
the fractal dimension $D_q$ as, $\tau_q = D_q(q-1)$ 
\cite{geisel1,moessner,Meo,lfsantos}. The fractal dimension can equivalently 
be computed from~\cite{lfsantos,bogomolny,fradkin} 
\begin{equation} D_q=\lim_{S\rightarrow \infty} \frac{S_q^{\nu}}{\log (2S+1)}, 
\quad S_q^{\nu}=-\sum_{m=1}^{2S+1} \frac{|\chi_{\nu}(m)|^{2q}}{(q-1)}.
\end{equation}
$S_q^{\nu}$ is the R\'enyi entropy corresponding to the $\nu$-th eigenvector 
$|\chi_{\nu}\rangle$. In Figs.\ \ref{tq_vs_q} (a) and (b) we have plotted 
$\log I_q$ and the average R\'enyi entropy $S_q=\sum_{\nu} S_q^{\nu}/(2S+1)$ 
respectively as a function of $\log (2S+1)$ for different values of $q$. From 
the slope of the linear fitting as shown in Fig.\ \ref{tq_vs_q} (a) we obtain 
$\tau_q$ which is plotted versus $q$ in Fig.\ \ref{tq_vs_q} (c) exhibiting a 
nontrivial behavior which confirms the existence of fractality in the 
eigenvectors. The fractal dimension is obtained from the slope of the dashed 
line at large $q$, $D_q \sim 0.6$ as shown in Fig.\ \ref{tq_vs_q} (c). These 
plots have been generated after a sufficiently large number of Fibonacci 
steps, say, $m \sim 30$, after which the behavior does does not change with 
$m$ as shown in Fig.\ \ref{tq_vs_q}. We also observed that the qualitative 
behavior remains the same for 
different choices of the driving parameters $\lambda$ and $T_0$. 

Further, we explored the fractality in the eigenspectrum by computing the 
local number of eigenstates $\Delta N_{\varepsilon}$ within the interval 
$\Delta \varepsilon$ around the eigenphase $\varepsilon$. Typically $\Delta N_{\varepsilon}$ 
follows the relation $\Delta N_{\varepsilon} \sim (\Delta \varepsilon)^{\alpha}$ as depicted in the $\log-\log$ plot in Fig.\ \ref{tq_vs_q} (d) for $\varepsilon \sim 0$, i.e., at the center of the band. From the slope of the linear fitting in Fig.\ \ref{tq_vs_q} (d) we obtain $\alpha \sim 0.69$ which signifies 
fractality in the Floquet spectrum~\cite{kohmoto,geisel1,geisel,pandit}.

\section{Effects of interactions}
\label{interaction}

\begin{figure}[ht]
\centering
\includegraphics[scale=0.17]{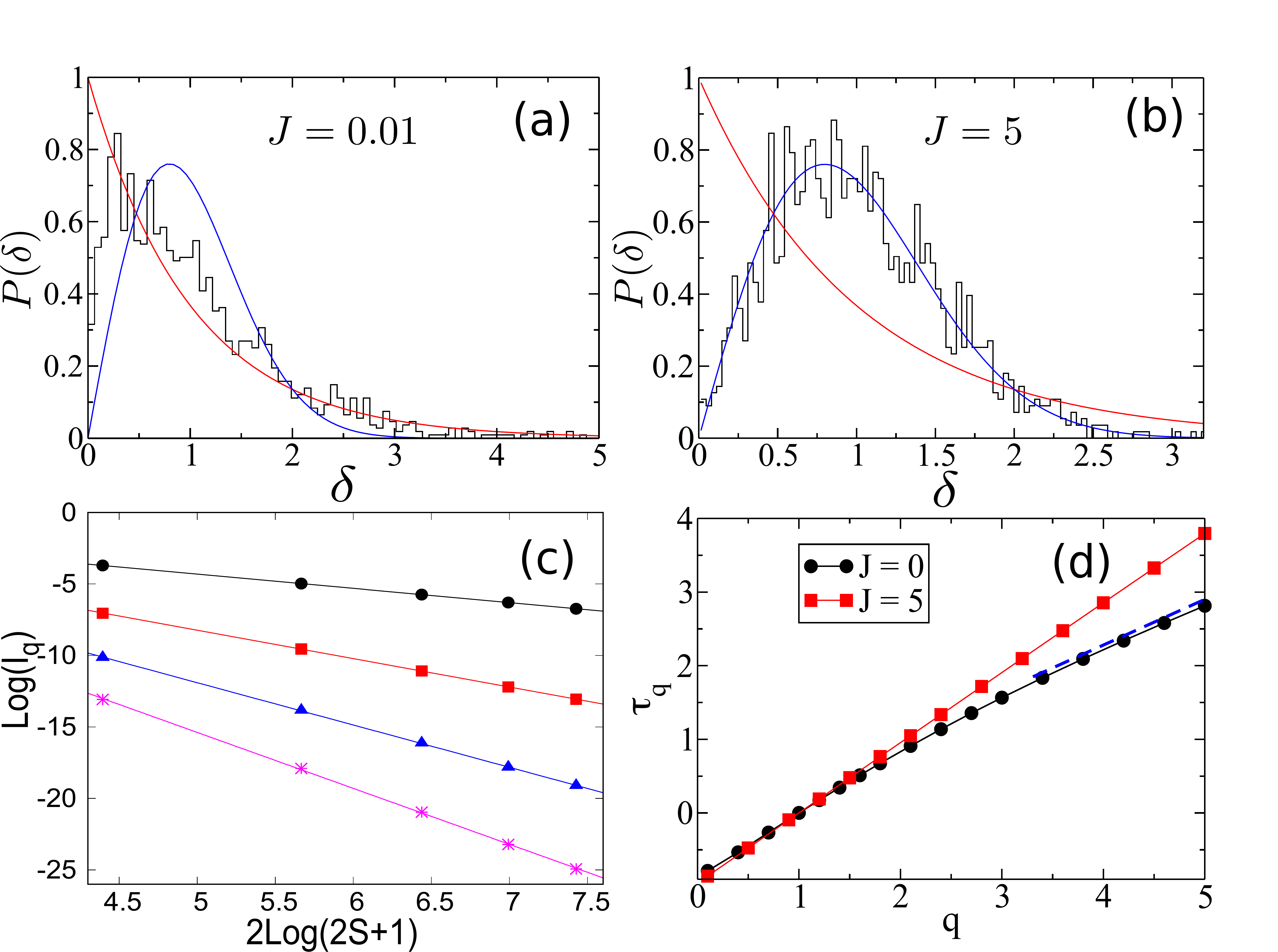}
\caption{(a)-(b) Spacing distribution $P(\epsilon)$ of eigenphases $\epsilon$ 
for $J = 0.01$ and $J = 5$ respectively, for $S = 20$ and computed at Fibonacci
step $n = 50$. (c) $\log I_q$ vs $2\log(2S+1)$ and (d) $\tau_q$ vs $q$ for 
$J = 5$ (red squares). The behavior of the corresponding non-interacting 
system ($J = 0$) is shown by the black circles. Here and in rest of the 
figures we set $T = \pi/100$ and $\lambda = \pi/10$.} \label{lev} \end{figure}

To study the effects of interactions, we consider a system of two interacting 
spins and the $\delta$-function kicking is applied to both the
spins following a Fibonacci sequence. Our goal is to study the effect of 
interactions on the quantum dynamics governed by the quasiperiodic drive. The 
Hamiltonian describing this system is given by
\begin{eqnarray} \hat{H}(t) &=& \hat{H}_0 ~+~ \lambda \hat{S}_x^{A/B} 
\sum_{n=-\infty}^{\infty} \delta \left(t-\sum_n T_n\right), \nonumber \\
\hat{H}_0 &=& \hat{S}_z^A ~+~ \hat{S}_z^B ~-~ J\hat{S}_z^A\hat{S}_z^B.
\end{eqnarray}
This is a simple generalization of the non-interacting model discussed above, 
where $J$ is the strength of the interaction between the two spins, and the 
last term represents kicking applied to the two spins following a Fibonacci 
sequence. The Floquet operators for the first two Fibonacci step are 
therefore given by
\begin{equation} \hat{\mathcal{F}}_{1}^{I} ~=~ e^{-iT\hat{H}_0} ~
e^{-i\lambda \hat{S}_x^A}, \quad \hat{\mathcal{F}}_{2}^{I} ~=~ 
e^{-iT\hat{H}_0} ~ e^{-i\lambda \hat{S}_x^B}. \end{equation}
The subsequent matrices in the Fibonacci sequence are then generated using 
Eq.\ \eqref{fibo_rec}. 
By diagonalizing the Floquet matrix $\hat{\mathcal{F}}_{m}^{I}$ we obtain the 
eigenphases $\varepsilon_{\nu}$ and the corresponding eigenvectors 
$|\chi_{\nu}\rangle$ (see Sec.\ \ref{qdyn}). We first compute the spacing 
between the successive eigenphases, i.e., $\delta_{\nu} = \varepsilon_{\nu+1}-
\varepsilon_{\nu}$, and calculate the distribution of the $\delta_{\nu}$'s 
following the procedure described in Ref.~\cite{Haake} in order to keep the 
normalization $\int P(\delta) d\delta = 1$ and mean $\int \delta P(\delta) d\delta = 1$. In Figs.\ \ref{lev} (a) and (b) we have plotted the normalized quasienergy spacing distribution $P(\delta)$ for small and large values of $J$ respectively.
We note that in the presence of interactions, the fractality of the system vanishes and level repulsion sets in. For large $J$, $P(\epsilon)$ resembles the Wigner surmise and corresponds to the Gaussian orthogonal ensemble of random 
matrix theory as shown in Fig.\ \ref{lev} (b); this indicates a possible 
thermalization of the system for large values of the interaction 
strength $J$~\cite{sray}.

We compute the moments of the Floquet eigenstates using Eq.\ \eqref{Iq_moment} 
followed by a calculation of the exponent $\tau_q$ which is related to 
the fractal dimension $D_q$. In Fig.\ \ref{lev} (c) we have plotted $\log I_q$ 
vs $2\log (2S+1)$ for different values of $q$. From the linear fitting we 
obtain the slope $\tau_q$ which we have plotted in Fig.\ \ref{lev} (b) as a 
function of $q$ and compared with the case $J=0$. We note that in contrast to 
the non-interacting case, we find for finite values of the interaction 
strength $J$ that $\tau_q \sim q$; this implies $D_q \sim 1$ indicating an 
ergodic nature of the Floquet eigenstates.

\begin{figure}[ht]
\centering
\includegraphics[scale=0.165]{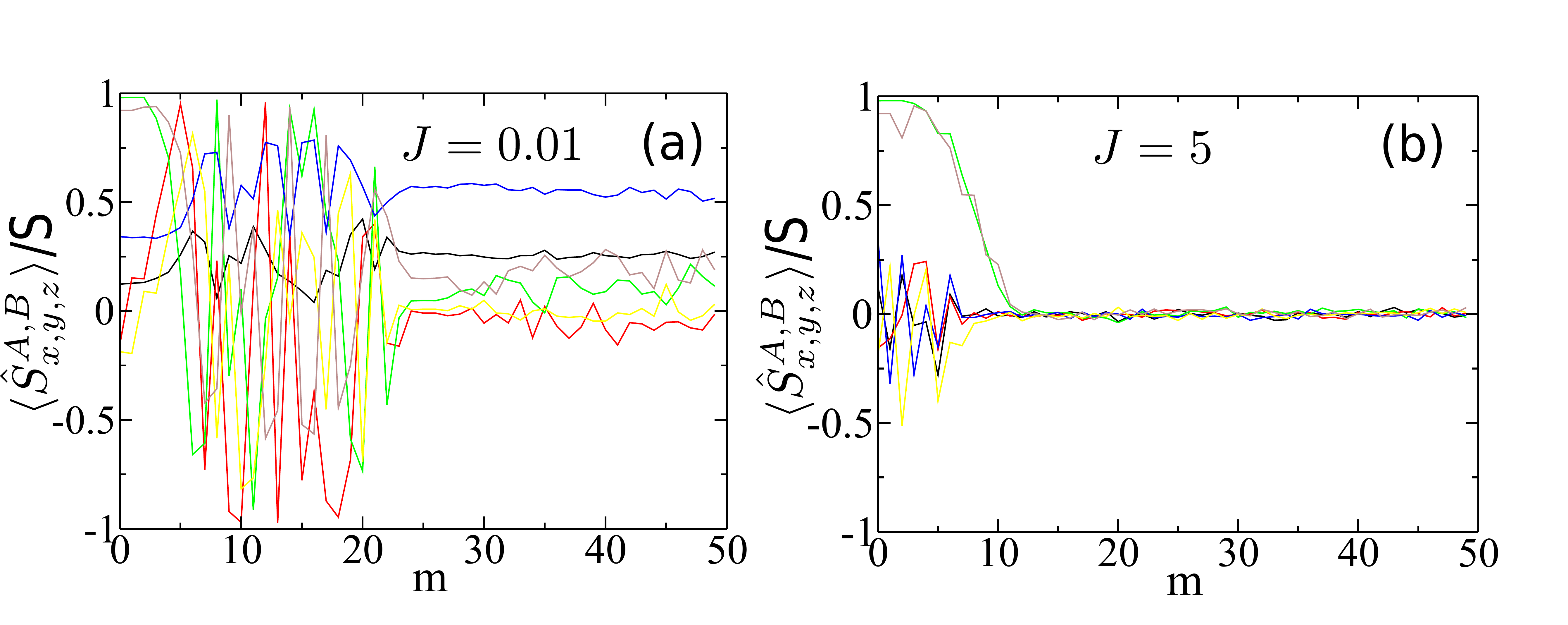}
\caption{(a)-(b) Time evolution of $\langle \hat{S}_{x,y,z}^{A,B}\rangle$ for 
$J = 0.01$ and $J = 5$ respectively, for $S = 10$.} \label{Si} \end{figure}

\begin{figure}[ht]
\centering
\includegraphics[scale=0.18]{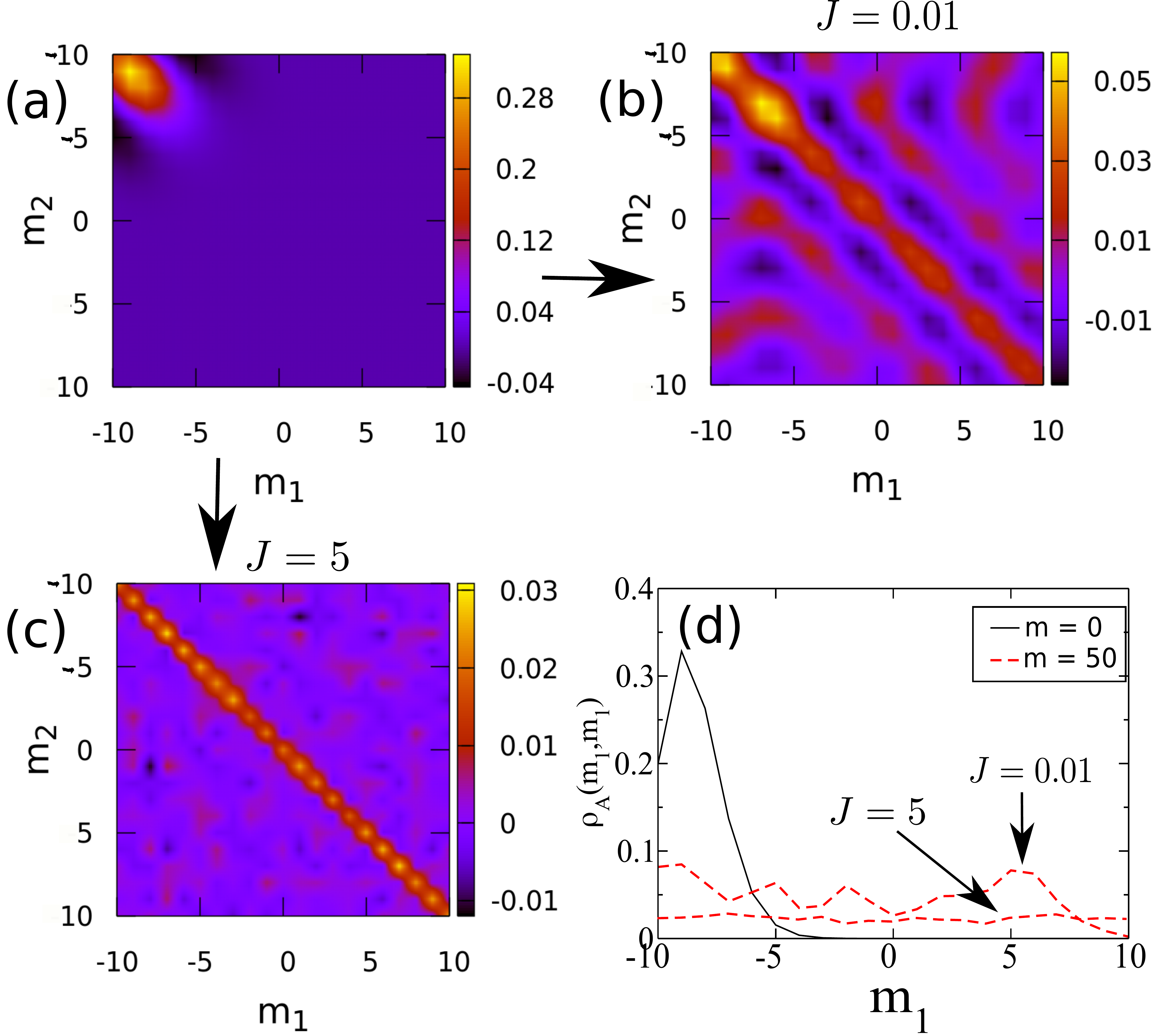}
\caption{(a)-(c) Time evolution of $\hat{\rho}_A$ for $J = 0.01$ and $J = 5$ 
respectively, for $S = 10$. (d) The corresponding diagonal elements are 
plotted.} \label{den} \end{figure}

We further elucidate this fact from the wave packet dynamics. We construct an 
initial wave function from the product of two spin coherent states given by
\begin{equation} |\psi_{AB}(0)\rangle ~=~ |\Theta, \Phi\rangle_{A} \otimes 
|\Theta, \Phi\rangle_{B}. \end{equation}
The expectation values of the corresponding spin observables 
$\langle \hat{S}_{x,y,z}^{A,B}\rangle$ can be computed from the time-evolved wave function $|\psi_{AB}(m)\rangle = \hat{\mathcal{F}}_m^{I} |\psi_{AB}(0)\rangle$. We observe that for small values of the interaction strength $J$, $\langle \hat{S}_{x,y,z}^{A,B}\rangle$ saturates to different nonzero values, whereas for 
large $J$, all the spin observables decays to zero as depicted in 
Figs.\ \ref{Si} (a) and (b) respectively. We compute the reduced density 
matrix corresponding to either of the spins A and B from the relation
\begin{equation} \hat{\rho}_{A(B)}^m ~=~ \text{Tr}_\text{B(A)}|\psi_{AB}(m)
\rangle \langle \psi_{AB}(m)|, \end{equation}
where $\text{Tr(.)}$ represents partial tracing with respect to spin B or A. 
In Fig.\ \ref{den} we have shown the structure of $\hat{\rho}_{A}^m$ after 
Fibonacci step $m = 50$ which, we have checked, is sufficient to obtain the 
steady states. For small values of $J$, $\hat{\rho}_{A}^m$ contains both the 
diagonal and off-diagonal entries in the eigenbasis of $\hat{S}_z$ indexed by 
$m_1$, $m_2$ which can be observed in Fig.\ \ref{den} (b). On the other hand, 
for large values of $J$, $\hat{\rho}_{A}^{m}$ becomes completely diagonal as 
is evident from Fig.\ \ref{den} (c), with equally weighted entries, i.e., 
$\rho_{A}^{m}(m_1,m_1) \sim 1/(2S+1)$. 
Such an observation indicates that in the presence of interactions the coherent
state picture is lost and leads us to conclude that the system approaches a
diagonal ensemble and thermalizes to infinite 
temperature~\cite{sray,fazio,nandkishore}.

\section{Conclusion}
\label{conclu}

To summarize, we have studied the dynamics of a spin-$S$ object which has a 
well-defined classical limit and is subjected to quasiperiodic kicking 
following the Fibonacci sequence. 
By evolving the corresponding classical Hamiltonian map of the spin variables 
we obtained the phase portraits which, for small kicking strength $\lambda$ and time period $T_0$, exhibits regular orbits which precess over an unit sphere. 
Interestingly, for increasing $\lambda$ and $T_0$ the dynamics appears to be 
chaotic; however, the Lyapunov exponent vanishes. 
We have calculated the Sutherland invariant which constrains to some
extent the dynamics governed by the transfer matrix with $SO(3)$ symmetry.
Fluctuations of the classical counterpart of the spin dynamics exhibit a 
fractal structure which is verified by its Fourier spectrum analysis.
It turns out that for an initially chosen spin coherent state, the 
phase coherence is retained during the time evolution under Fibonacci drive 
even for large $\lambda$ and $T_0$ indicating q classical-quantum 
correspondence. Fractality in classical dynamics is observed from the 
spectral analysis.
More interestingly, the fractality is also present in the internal structure of the Floquet matrix governing the full quantum dynamics which has been 
investigated from the scaling of the R\'enyi entropy, as well as from 
the moments of the Floquet eigenvectors and the quasienergy spectrum.
Finally, we have considered two such spin-$S$ objects interacting with each 
other and driven quasiperiodically. We have shown that in the presence of 
the interaction, the fractal behavior vanishes and level repulsion sets in 
the Floquet quasienergy spectrum. 
In the dynamics, we observed that for increasing interaction strength $J$, 
the average values of the components of spin operators for both the spins saturate to zero and 
the reduced density matrix for either of the spins becomes diagonal. The 
emergence of the diagonal ensemble with equally weighted diagonal elements 
indicates thermalization of the system to infinite temperature and 
corresponds to the microcanonical ensemble of statistical mechanics.
\\

In conclusion, the existence of critical Floquet eigenstates exhibiting fractality in a quasiperiodically driven non-interacting spin model and its crossover 
to ergodic dynamics in the presence of interactions constitutes the central 
result of our work. Kicked spin models have already been realized in 
experiments considering the angular momentum of an atom in a suitable hyperfine state~\cite{q_state_exp,q_state_exp1}; 
the kicking can be generated by using a short magnetic pulse~\cite{k_top}. The
dynamics of such kicked systems can also be investigated in circuit QED 
experiments~\cite{Cqed}. The signature of fractality and its change to ergodic 
behavior in the dynamics can be found in the experiments by measuring the 
discrete time Fourier amplitude spectrum of time varying physical observable such as 
the spin variables in our study as well from the dimension 
measurement~\cite{SNA_exp}. The models discussed in our paper can thus be 
realized experimentally and our results can be tested in similar experiments.

\section*{Acknowledgments} 

S.R. thanks Anandamohan Ghosh for useful discussions. D.S. thanks DST, India 
for Project No. SR/S2/JCB-44/2010 for financial support. S.R. acknowledges the
financial support and hospitality provided by IISER Kolkata for this work.


\begin{thebibliography}{99}
\bibitem{Floquet} Here we call the time evolution operator between two consecutive kickings as the Floquet operator.
\end{thebibliography}
\end{document}